\documentclass[11pt, a4paper]{article}

\usepackage{jheppub}

\usepackage{mathrsfs}
\usepackage[T1]{fontenc}
\usepackage{mathpazo}
\usepackage{setspace}
\usepackage{amsfonts}
\usepackage{amssymb}
\usepackage{amsmath}
\usepackage{epsfig}
\usepackage{latexsym}
\usepackage{color}
\usepackage{graphicx}
\usepackage{nicefrac}
\usepackage[latin1]{inputenc}
\usepackage{pstricks}
\usepackage{slashed}
\usepackage{multirow}
\usepackage{hyperref}

\def\be{\begin{equation}}
\def\ee{\end{equation}}
\def\ba{\begin{eqnarray}}
\def\ea{\end{eqnarray}}

\def\del{\partial}

\def\s{\sigma}
\def\S{\Sigma}

\def\IR{\relax{\rm I\kern-.18em R}}

\def\inv{^{\raise.0ex\hbox{${\scriptscriptstyle -}$}\kern-.05em 1}}

\title{non-Abelian T-duality, $G_2$-structure Rotation and Holographic Duals of $\mathcal{N}=1$ Chern-Simons Theories}

\author[a]{Niall T. Macpherson,}
 \affiliation[a]{Department of Physics, Swansea University, Singleton Park, Swansea, SA2 8PP, UK \\}
 \emailAdd{pymacpherson@swansea.ac.uk}

\abstract{A new dynamic $SU(3)$-structure solution in type-IIA is found by T-dualising a deformation of the Maldacena-Nastase solution along an $SU(2)$ isometry. It is argued that this is dual to a quiver gauge theory with multiple Chern-Simons levels. A clear way of defining Chern-Simons levels in terms of Page charges is presented, which is also used to define a Chern-Simons term for the $G_2$-structure analogue of Klebanov-Strassler, providing evidence of a cascade in both the ranks and levels of the dual quiver.}

\def\beq{\begin{equation}}
\def\eeq{\end{equation}}
\def\bea{\begin{eqnarray}}
\def\eea{\end{eqnarrat}}
\begin{document}

\maketitle
\flushbottom

\section{Introduction}
Almost from its conception, supergravity solution generating techniques have played an important role in the gauge/gravity correspondence. An early example is the method by which one can lift a lower dimensional gauged supergravity to a full 
solution of type-II supergravity in 10-d. In \cite{Schvellinger:2001ib} such a lift was performed on  the 5-d solution of \cite{Chamseddine:2001hk}. The resulting 10-d solution was identified as a holographic dual of $\mathcal{N}=1$ SYM in 3-d with gauge group $SU(N_c)$ and Chern-Simons level $k=N_c/2$ by Maldacena and Nastase \cite{Maldacena:2001pb}, after whom the solution is usually named.  

The Maldacena-Nastase solution consists of wrapped D5-branes wrapping a 3-cycle in a manifold that supports a $G_2$-structure. The field theory living on the world volume of these branes is only effectively 3-d in the IR and so the geometry is only a good description of the low energy dynamics of SYM in 3-d. A UV completion is provided by another solution generating technique, $G_2$-structure rotation \cite{Gaillard:2010gy}, which is analogous to U-duality. This acts on the S-dual of a deformation of the Maldacena and Nastase solution \cite{Canoura:2008at} and maps it to a geometry supporting D2 and fractional D2 branes that asymptotes to $AdS_4\times Y$ in the UV and is the $G_2$ analogue of the baryonic branch of Klebanov-Strassler \cite{Klebanov:2000hb}. The compact metric $Y$ has finite volume in the UV and so the fractional branes which wrap cycles in $Y$ remain effectively 3-d in the whole geometry. A gauge theory analysis of the $G_2$-structure rotated solution was performed in \cite{Macpherson:2013gh} which suggested that the dual QFT was likely a confining 2 node quiver with a Chern-Simons term that dominates the IR dynamics and a conformal fixed point in the UV. This clearly presents an improvement on the UV behaviour of the geometry and coupled with the possibility of a duality cascade by analogy with Klebanov-Strassler, indicates that the dual field theory is potentially very interesting.

The main focus of this work will be to apply a different solution generating technique, non-abelian T-duality, to the deformed Maldacena and Nastase solution. Dualising supergravity solutions along non-abelian isometries is actually quite an old idea \cite{delaossa:1992vc}. However it is only quite recently that it was realised how to dualise geometries supporting a non-trivial RR sector in \cite{Sfetsos:2010uq} with refinements given by \cite{Lozano:2011kb,Itsios:2012dc}. This was of course a necessary step before the duality could have much utility in the gauge/gravity correspondence. Most attention has been focused on dualising along $SU(2)$ isometries, because they are quite simple and it has been explicitly shown (for a quite general ansatz) that they are always a map between SuGra solutions \cite{Jeong:2013jfc}. This has already bore some quite interesting results, for example a new $AdS_6$ solution was generated in type-IIB \cite{Lozano:2012au}, which promises to shed some light on CFT'S in 5-d. Attention has also been focused on performing $SU(2)$ T-dualities on type-IIB conifold solutions \cite{Itsios:2012zv,Itsios:2013wd}, with the aim of generating new type-IIA solutions that give holographic descriptions of confining 4-d gauge theories. The supersymmetric solution on the conifold  support an $SU(3)$-structure and it was shown in \cite{Barranco:2013fza} (see also \cite{Macpherson:2013lja}) that their non-abelian T-dual geometries must support an $SU(2)$-structure if SUSY is preserved. This was an important step towards understanding the duality in terms of generalised geometry \cite{Grana:2005sn}, which has been of considerable help in the context of holography and gives an explicit check of SUSY preservation by the duality. In addition to this though, $SU(2)$-structure backgrounds are not that common and so a method of generating them should led to greater insight of their general construction.\\

In this work a new type-IIA SuGra solution, dual to a confining gauge theory in 3-d, is generated by performing a T-duality along one of the $SU(2)$ isometries of the deformed Maldacena-Nastase solution  \cite{Canoura:2008at}. This new solution preserves $\mathcal{N}=1$ SUSY in the form of a dynamic $SU(3)$-structure in 7-d \cite{Gauntlett:2002sc,Martelli:2003ki}. Some details of the $G_2$-structure rotation \cite{Gaillard:2010gy} are also given so that the two solutions may be compared. This also includes a new proposal for the Chern-Simons term of the $G_2$-structure rotation in term of a probe D8 brane with the level given by the D6 Page charge, such a proposal is not present in literature. The gauge theory dual to the T-dual geometry is analysed and compared to that of the original wrapped D5 brane and $G_2$-structure rotated solutions. A more specific outline is as follows:

In section \ref{Sec: II} the deformation of the Maldacena-Nastase solution is briefly reviewed. Details of the metric and RR sector ansatz and $G_2$-structure SUSY equations are all presented. SUSY preserving semi-analytic solutions of the ansatz are given that are characterised by either a UV constant or UV linear dilaton. And finally some cycles and charges that will be of relevance to the field theoretic description are introduced.

Section \ref{Sec: III} gives some details of the solution generating technique $G_2$-structure rotation and  presents the result of applying this to the deformed Maldacena-Nastase solution. Some details of the rotated $G_2$-structure SUSY conditions are given and cycles and Page charges, that will be of interest in penultimate section, are introduced.

Section \ref{sec: IV} is where the new results begin, the reader familiar with the salient features of the Maldacena-Nastase solution and its $G_2$-structure rotation may wish to start here. The section begins with a brief review of non-abelian T-duality on $SU(2)$-isometries before the dual geometry is presented in as concise way as possible. After this attention is turned to the generalised geometric description of the dual soluition. It is shown that the dual structure is dynamic $SU(3)$ in 7-d which is characterised by a non-constant angle between the two 10-d MW Killing spinors of  type-IIA. Finally some cycles and Page charges are introduced that will be important in the field theory analysis. 

Section \ref{sec: probes} contains a field theory analysis of each of the solutions presented in the previous sections. The analysis of the deformed Maldacena-Nastase and its $G_2$-structure rotation is mostly a review of what can be found in \cite{Maldacena:2001pb,Macpherson:2013gh,Canoura:2008at,Macpherson:2012zy} although additional clarifications are made. In particular further details of a Seiberg like duality in the $G_2$-structure rotated solution are given and how this effects the Chern-Simons level, the proposal for which is new. The analysis of the T-dual geometry suggests that it is dual to a confining Chern-Simons like gauge theory that is potentially a 3-node quiver.

The work is finally closed with concluding remarks and an outlook in section \ref{sec: conclusion}

\section{Wrapped D5 Branes on $\Sigma^3$}\label{Sec: II}
The Maldacena-Nastase solution \cite{Maldacena:2001pb} is a solution of type-IIB (first presented in \cite{Schvellinger:2001ib}) that consists of D5 branes wrapping a 3-cycle in a $G_2$-structure manifold and is dual in the IR to $\mathcal{N}=1$ SYM in 3-d. The purpose of this section is to review its deformation due to Canoura, Merlatti and Ramallo \cite{Canoura:2008at}, as this constitutes more general ansatz to a set of wrapped D5 brane solutions \cite{Canoura:2008at,Gaillard:2010gy,Macpherson:2012zy,Macpherson:2013gh,Macpherson:2013dta} which contain the Maldacena-Nastase solution as a special case \footnote{Actually further modification of the RR 3-form is required to include sources, which is the main focus of the majority of these references.}. 

The string frame metric is given by:
\begin{equation}\label{met}
ds^2_{\text{str}} = e^{\phi}\bigg(dx^2_{1,2}+ds^2_7\bigg)
\end{equation}
where the internal part of the metric, $ds^2_7$, describes a manifold supporting a $G_2$-structure and is given by
\begin{equation}\label{Hand7dmet}
ds^2_7=  N_c \bigg[e^{2g}dr^2+\frac{e^{2h}}{4}(\sigma^i)^2+\frac{e^{2g}}{4}(\omega^i-\frac{1}{2}(1+w)\sigma^i)^2\bigg]\\
\end{equation}
The functions $g$, $h$, $w$ and the dilaton $\phi$ all depend on the holographic coordinate $r$ only. $\sigma^i$ and $\omega^i$ are 2 sets of $SU(2)$ left invariant 1-forms which satisfy the following differential relations:
\begin{equation}
d\sigma^i=-\frac{1}{2}\epsilon_{ijk}\sigma^j\wedge\sigma^k;~~~~~d\omega^i=-\frac{1}{2}\epsilon_{ijk}\omega^j\wedge\omega^k
\end{equation}
These can be represented by introducing 3 angles for $\sigma^i$, $(\theta_1,\phi_1,\psi_1)$ and a further 3 for $\omega^i$, $(\theta_2,\phi_2,\psi_2)$ such that:
\begin{equation}\label{rep}
\left.\begin{array}{l l}
\vspace{3 mm}
\sigma^1=&\cos\psi_1 d\theta_1 +\sin\psi_1\sin\theta_1 d\phi_1\\
\vspace{3 mm}
\sigma^2=&-\sin\psi_1 d\theta_1 +\cos\psi_1\sin\theta_1 d\phi_1\\
\vspace{3 mm}
\sigma^3=& d\psi_1+\cos\theta_1 d\phi_1
\end{array}\right.
\end{equation}
and similarly for $\omega^i$. The angles are defined over the ranges: $0\leq\theta_{1,2}\leq\pi$, $0\leq\phi_{1,2}<2\pi$ and $0\leq\psi_{1,2}<4\pi$.
The solution has a non-trivial RR 3-form:
\begin{equation}\label{F3}
\left.\begin{array}{l l}
F_3=&\frac{N_c}{4}\bigg[\left(\sigma^1\wedge\sigma^2\wedge\sigma^3-\omega^1\wedge\omega^2\wedge\omega^3\right)\! +\frac{\gamma'}{2}dr\wedge\sigma^i\wedge\omega^i +\\&
~~~~~~\frac{1}{4}\epsilon_{ijk}\left(\left(1\!+\!\gamma\right)\sigma^i\wedge\sigma^j\wedge\omega^k\!-\!\left(1\!+\!\gamma\right)\omega^i\wedge\omega^j\wedge\sigma^k\right)\bigg]
\end{array}\right.
\end{equation}
which satisfies the simple Bianchi identity 
\beq
dF_3=0
\eeq
This solution preserves $\mathcal{N}=1$ SUSY in 3-d, which is 2 real supercharges. This can be expressed in terms of the following differential constraints on an associative 3-form $\Phi_3$: 
\beq
\begin{array}{ll}\label{eq: Gstr}
&d\big(e^{2A-\phi}\big)=0\\
&\Phi_3\wedge d\Phi_3=0\\
&d\big(e^{2A-\phi}\star_7\Phi_3\big)=0\\
&d\big(e^{3A-\phi}\Phi_3\big)+e^{3A}\star_7 F_3=0
\end{array}
\eeq
where $A=\phi/2$. Generically the 3-form $\Phi_3$ may be expressed in terms of an auxiliary $SU(3)$-structure as \cite{Fwitt}:
\beq\label{eq: 3form}
\Phi_3=e^r\wedge J + Re \Omega_{hol},~~~~\star_7\Phi_3=\frac{1}{2}J\wedge J + Im \Omega_{hol}\wedge e^r
\eeq
a convenient set of vielbeins for the metric are given by
\beq\label{eq: orgviels}
\begin{array}{l l}
\vspace{3 mm}
&e^{x_i}=e^{\phi/2}dx_i,~~~~e^{r}=\sqrt{N_c}e^{\phi/2+g}dr,~~~~ e^{i}=\sqrt{N_c}\frac{e^{\phi/2+h}}{2}\sigma^i,\\
&e^{\hat{i}}=\sqrt{N_c}\frac{e^{\phi/2+g}}{2}\big(\omega^i-\frac{1}{2}(1+w)\sigma^i\big)\\
\end{array}
\eeq
with respect to which auxiliary $SU(3)$-structure of the deformed Maldacena-Nastase background can be expressed as \cite{Haack:2009jg,Gaillard:2010gy}:
\beq\label{eq: SU3str}
\begin{array}{ll}
\vspace{3 mm}
&J= e^{\hat{1}}\wedge e^{1}+e^{\hat{2}}\wedge e^{2} + e^{\hat{3}}\wedge e^{3} \\
&\Omega_{hol}= i e^{i \alpha}(e^{\hat{1}}+i e^{1})\wedge(e^{\hat{2}}+i e^{2})\wedge(e^{\hat{3}}+i e^{3})
\end{array}
\eeq
where $\alpha$ depends on r. Plugging eq \ref{eq: SU3str} into eq \ref{eq: Gstr} gives rise to a set of first order differential equations that are presented, for example, in appendix A of \cite{Macpherson:2013gh}\footnote{There a flavour brane profile $P(r)$ is also considered which should be set to zero when there are no sources.}. The solution of these equations is only known semi analytically in the IR and UV, but it is possible to numerically interpolate between these two sets of series solutions. 

In the IR where $r\sim 0$ the solution is regular and is given by
\beq
\begin{split}
e^{2g}&=g_0+\frac{(g_0-1)(9g_0+5)}{12g_0}r^2+O(r^4),\\
e^{2h}&=g_0 r^2-\frac{3g_0^2-4g_0+4}{18g_0}r^4+O(r^6),\\
w&=1-\frac{3g_0-2}{3g_0}r^2+O(r^4),\\
\gamma&=1-\frac{1}{3}r^2+O(r^4),\\
\phi&=\phi_0+ \frac{7}{24g_0^2}r^2.
\end{split}
\eeq
Notice that $g_0=1$ seems to be special, the solution to $e^{2g}$  truncates and $\gamma=w$, indeed this persists to all orders in $r$. This is the Maldacena-Nastase solution, its UV expansion about $r\sim\infty$ is characterised by an asymptotically linear dilaton
\beq\label{eq: UV1}
\begin{split}
e^{2g}&=1,\\
e^{2h}&=2r + h_0+\frac{1}{8r}+\frac{1-2 h_0}{32 r^2}+O(r^{-3}),\\
w&=\frac{1}{4r}+\frac{5-4 h_0}{32r^2}+O(r^{-3}),\\
\gamma&=w,\\
\phi&=\phi_{\infty}+r -\frac{3}{8}\log r -\frac{3h_0}{16 r}+O(r^{-2}).
\end{split}
\eeq
where the constant needs to be fine tuned to the value $h_0=-\frac{3}{2}$ so that the IR and UV numerically matched. 

When $g_0>1$ the solution is a deformation of Maldacena-Nastase characterised by an asymptotically constant dilaton
\beq\label{eq: UV2}
\begin{split}
e^{2g}&=c e^{4r/3} -1 +\frac{33}{4c}e^{-4r/3}+O(e^{-8r/3}),\\
e^{2h}&=\frac{3c}{4}e^{4r/3}+\frac{9}{4}-\frac{77}{16c}e^{-4r/3}+O(e^{-8r/3}),\\
w&=\frac{2}{c}e^{-4r/3}+O(e^{-8r/3}),\\
\gamma&=\frac{1}{3}+O(e^{-8r/3}),\\
\phi&=\phi_{\infty}+\frac{2}{c^2}e^{-8r/3}+O(e^{-2r}).
\end{split}
\eeq
at higher orders polynomial terms also appear and a sub expansion in odd powers of $e^{-2r/3}$ has been set to zero \cite{Macpherson:2013gh}. The UV constant $c$ must be tuned for specific choices of the IR constant $g_0$ such that the series solutions may be smoothly connected numerically.

It is possible to show that the flux equation of motion
\beq
d\star F_3=0
\eeq
is satisfied once eq \ref{eq: Gstr} is imposed and likewise are Einstein's equations and the dilaton equation of motion.
The last line of eq \ref{eq: Gstr} gives a definition of the potential $C_6$ such that $dC_6=F_7$:
\beq
C_6= e^{3A}Vol_3\wedge \Phi_3
\eeq

There are several important 3-cycles in the geometry which are related to gauge theory observables that shall be discussed at length in section \ref{sec: probes} , these are:
\beq\label{eq: Mna cycles}
S^3=\{\sigma^{i}| \omega^i=0\},~~~~~ \tilde{S}^3=\{\omega^{i}| \sigma^i=0\},~~~~~\Sigma^3= \{\sigma^{i}=\omega^i\}.
\eeq
The induced metrics on $S^3$ and $\tilde{S}^3$ are non-vanishing in the whole geometry and are thus suitable for defining flux quantisation. The integrals of $F_3$ on these cycles give respectively
\beq
\int_{S^3} F_3=-\int_{\tilde{S^3}} F_3=2\kappa_{10} T_5 N_c .
\eeq
once one sets $2\kappa_{10}=(2\pi)^7$ and $T_p=\frac{1}{(2\pi)^p}$.
The pullback of $F_3$ onto $\Sigma^3$ is zero, while the induced metric
\beq
ds^2_{\Sigma^3}=\frac{e^{\phi}N_c}{4}\bigg[e^{2h}+\frac{e^{2g}}{4}(w-1)^2\bigg](\sigma^i)^2
\eeq
vanishes in the IR and blows up in the UV. This is the 3-cycle on which the D5 branes are wrapped, their world volume becomes 3-dimensional in the IR as the cycle shrinks to zero but the background remains non-singular because $F_3$ vanishes on $\Sigma^3$.
\section{Solution Generating Technique I: $G_2$-Structure Rotation}\label{Sec: III}
In \cite{Gaillard:2010gy} a solution generating technique was found by Gaillard and Martelli that maps any unwarped type-IIA $G_2$-structure solution with asymptotically constant dilaton and NS 3-form flux H to a more exotic $G_2$-structure solution with a non-trivial RR sector. This method of solution generating is referred to as Rotation, as it acts on the space of Killing spinors thus, but can also be viewed as a U-duality\footnote{for closely related work on $SU(3)$-structure rotations in IIB see, for example, \cite{Gaillard:2010qg,Elander:2011mh,Bennett:2011va,Conde:2011aa} and for U-duality \cite{Maldacena:2009mw,Caceres:2011zn}}.

If one dimensionally reduces the M-theory solution of \cite{Martelli:2003ki} one is left with a $G_2$-structure solution in type-IIA. Its string frame metric is:
\beq
ds^2_{str}=e^{2\Delta+2\hat{\phi}/3}(dx^2_{1,2}+d\hat{s}^2_7)
\eeq
where $\hat{\phi}$ is the dilaton and $d\hat{s}^2_7$ is any $G_2$-structure manifold.
The condition that $\mathcal{N}=1$ SUSY is preserved can be expressed as the following differential relations between the fluxes, the 3-form $\hat{\Phi}_3$ and a phase $\zeta$:
\beq\label{eq:RotBPS}
\begin{array}{ll}
\vspace{3 mm}
&\hat{\Phi}_3\wedge d\hat{\Phi}_3=0\\
\vspace{3 mm}
&d(e^{6\Delta} \star_7 \hat{\Phi}_3)=0\\
\vspace{3 mm}
&d(e^{2\Delta+2\hat{\phi}/3}\cos\zeta)=0\\
\vspace{3 mm}
&2d\zeta-e^{-3\Delta}\cos\zeta d(e^{3\Delta}\sin\zeta)=0\\
\vspace{3 mm}
&\frac{1}{\cos^2\zeta}e^{-4\Delta+2\hat{\phi}/3}\star_7 d(e^{6\Delta} \cos\zeta \hat{\Phi}_3)= H_3\\
\vspace{3 mm}
&Vol_3\wedge d(e^{3\Delta}\sin\zeta)-\frac{\sin\zeta}{\cos^2\zeta}e^{-3\Delta}d(e^{6\Delta} \cos \zeta\hat{\Phi}_3)=F_4\\
\end{array}
\eeq
The central observation of Gaillard and Martelli was that if one sets $\zeta =0$ eq \ref{eq:RotBPS} truncates to the S-dual of eq \ref{eq: Gstr}, that is:
\beq
\begin{array}{ll}
\vspace{3 mm}
&\Phi^{(0)}_3\wedge d\Phi^{(0)}_3=0\\
\vspace{3 mm}
&d(e^{-2\phi^{(0)}} \star_7 \Phi^{(0)}_3)=0\\
\vspace{3 mm}
&e^{2\phi^{(0)}} d(e^{-2\phi^{(0)}} \Phi^{(0)}_3)+\star_7H_3=0\\
\end{array}
\eeq 
and the metric is simply
\beq
ds^2_{(0)}= dx^2_{1,2}+ ds^{(0)2}_7.
\eeq
Any solution of this simplified system will also be a solution of eq \ref{eq:RotBPS} when the following identifications are made:
\beq
\begin{array}{ll}
\vspace{3 mm}
&~~~~~~~~~~~~~~~~~~~\hat{\Phi}_3=\left(\frac{\cos\zeta}{\kappa_1}\right)^3 \Phi^{(0)}_3,~~~~e^{2\hat{\phi}}=\frac{\cos\zeta}{\kappa_1}e^{2\phi^{(0)}}\\
&e^{3\Delta}=\left(\frac{\kappa_1}{\cos\zeta}\right)^2 e^{-\phi^{(0)}},~~~~
d\hat{s}^2_7=\left(\frac{\cos\zeta}{\kappa_1}\right)^2 ds^{(0)2}_7,~~~~\sin\zeta =\kappa_2e^{-\phi^{(0)}}
\end{array}
\eeq
where $\kappa_1$ and $\kappa_2$ are integration constants and $\phi^{(0)}$ must be bounded to satisfy the last equation.

It is possible to perform a rotation of the deformed Maldacena-Nastase solution, detailed in the last section, once an S-duality has been performed on it. This sends
\beq
F_3\to H_3,~~~~ \phi \to \phi^{0}=-\phi, ~~~~ ds^2_{str}\to e^{-\phi}ds^2_{str}
\eeq
so that the resulting metric is unwarped. Specifically it is the solution with UV given by eq \ref{eq: UV2} that is suitable for this as the dilaton is bounded. The 3-form, $\Phi_3^{(0)}$ is still given by eq \ref{eq: 3form} but with the auxiliary $SU(3)$-structure of eq \ref{eq: SU3str} with no dilaton factor
\beq
\hat{e}^a= e^{-\phi/2}e^{a}.
\eeq
As the solution is now in the common type-II NS sector it can be viewed as a type-IIA theory, as required by the rotation. The interested reader is referred to \cite{Gaillard:2010gy} for further  details of the solution generating algorithm. 

The rotated solution has a warped metric and modified dilaton, which after fixing the integration constants and rescaling the field theory coordinates may be expressed as:
\beq
\begin{array}{ll}
\vspace{3 mm}
ds^2_{str}&= \frac{1}{c \sqrt{H}}dx^2_{1,2}+\sqrt{H} ds^2_7\\
\vspace{3 mm}
e^{2\hat{\phi}}&=c\sqrt{H}e^{2(\phi^{(0)}-\phi_{\infty})}\\
H&=1-e^{-2(\phi^{(0)}-\phi_{\infty})}
\end{array}
\eeq
where $\hat{\phi}$ is the new dilaton, $\phi_{\infty}$ is the UV value of $\phi^{(0)}$ and $c$ is a constant which appears in the UV series solutions to the BPS equations \cite{Macpherson:2013gh}, but which will not play an important role here. The metric in string frame tends in the UV towards $AdS_4\times Y$ where $Y$ is the metric at the base of a $G_2$-cone, however the dilaton is not constant, $e^{2\hat{\phi}}\sim e^{-8r/3}$, and so the solution does not enjoy full conformal symmetry.

The NS 3-form is unchanged but an RR 4-form has been generated:
\beq
\left.\begin{array}{l l}
H_3=&\frac{N_c}{4}\bigg[\left(\sigma^1\wedge\sigma^2\wedge\sigma^3-\omega^1\wedge\omega^2\wedge\omega^3\right)\! +\frac{\gamma'}{2}dr\wedge\sigma^i\wedge\omega^i +\\
\vspace{3 mm}
&~~~~~~\frac{1}{4}\epsilon_{ijk}\left(\left(1\!+\!\gamma\right)\sigma^i\wedge\sigma^j\wedge\omega^k\!-\!\left(1\!+\!\gamma\right)\omega^i\wedge\omega^j\wedge\sigma^k\right)\bigg]\\
F_4=& -\frac{1}{c^2} Vol_3\wedge dH^{-1} +\frac{\sqrt{N_c}}{\sqrt{c}}e^{2(\phi_{\infty}-\phi^{(0)})}\star_7 H_3
\end{array}\right.
\eeq
these obey the Bianchi identities
\beq
dH_3=0,~~~~ dF_4=0
\eeq
and flux equations of motion
\beq
\begin{array}{ll}
\vspace{3 mm}
&d\star F_4+ H_3\wedge F_4=0\\
&d(e^{-2\hat{\phi}} \star H_3)-\frac{1}{2} F_4\wedge F_4 =0
\end{array}
\eeq
One can use eq \ref{eq:RotBPS} to define a canonical potential $C_3$ such that $dC_3=F_4$
\beq\label{eq: C3pot}
C_3= \frac{1}{c^2}Vol_3\wedge d H^{-1}+\frac{1}{\sqrt{c}}e^{2(\phi_{\infty}-\phi^{(0)})}\Phi_3^{(0)}.
\eeq

In \cite{Macpherson:2013gh} some cycles of interest were identified. Those that give flux quantisation are:
\beq
\hat{\Sigma}^3=\Sigma^3=\{\sigma^i=\omega^i\},~~~~\hat{\Sigma}^6=\{\s^1,\s^2,\s^3,\omega^1,\omega^2,\omega^3\}.
\eeq
The Maxwell and Page charges \cite{Marolf:2000cb} coincide for the NS5 brane (as they did for the D5 brane in the previous section). However the flux equation of motion for $F_4$ implies that this is not so for the D2 brane and it is only the Page charge that is quantised for this object. Define $F_6=-\star F_4$, then the following charges are quantised.
\beq\label{eq: NS5andD2charges}
\begin{split}
\vspace{3 mm}
Q_{NS5}&=-\frac{1}{4\pi^2}\int_{\tilde{S}^3}H_3=N_c\\
Q_{D2}&= -\frac{1}{(2\pi)^5}\int_{\Sigma^6}\big(F_6+H_3\wedge C_3\big)=0 ~mod ~N_c.
\end{split}
\eeq
Actually substituting eq \ref{eq: C3pot} into the definition of $Q_{D2}$ gives zero, but $C_3$ is not a gauge invariant quantity and this gives rise to non-zero integers under the large gauge transformation. Let $Q_{D2}=M_c$ then consider the large gauge transformation $C_3\to C_3 +\Delta C_3$ where
\beq\label{eq: gaugetrans}
\Delta C_3=-\frac{ \pi}{4}\bigg[\sigma^1\wedge\sigma^2\wedge\sigma^3+\omega^1\wedge\omega^2\wedge\omega^3\bigg].
\eeq
This will shift the Page charge as $Q_{D2}\to M_c- N_c$.

Another cycle with interesting properties is the 2-cycle at constant $r$:
\beq\label{eq: sigmahat2}
\hat{\Sigma}^2=\{\theta_1=\theta_2,\varphi_1=\varphi_2|\psi_1=\psi_2=\text{Constant}\}
\eeq
On this cycle $F_4$ vanish and the induced metric
\beq
ds^2_{\hat{\Sigma}^2}= \frac{N_c}{4}\sqrt{H}\big(e^{2h}+\frac{e^{2g}}{4}(w-1)^2\big)\big(d\theta_1^2+d\varphi_1^2\big),
\eeq
has vanishing volume in the IR and constant volume in the UV.
\section{Solution Generating Technique II: non-Abelian T-duality}\label{sec: IV}
It was relatively recently that a method of performing T-duality on non-abelian isometries was realised for backgrounds with non-zero RR fluxes \cite{Sfetsos:2010uq}. This has been the focus of some study in recent years \cite{Lozano:2011kb,Itsios:2012dc,Lozano:2012au,Itsios:2012zv,Itsios:2013wd,Barranco:2013fza} with the aim of utilising the duality as a SuGra solution generating technique in the context of the gauge-gravity correspondence. 

The purpose of this section is to present the first non-abelian T-dual of a background with minimal SUSY in 3-d, specifically a dual of deformed Maldacena-Nastase along an $SU(2)$ isometry. First however the duality procedure shall be briefly reviewed.
\subsection{A Concise Review}
A general prescription for performing an $SU(2)$ non-abelian T-duality is presented in \cite{Itsios:2013wd}, with more generic isometries considered in \cite{Lozano:2011kb}. Here the more salient details of the $SU(2)$ isometry case will be laid out.

A generic metric and NS 2-form of a background containing an $SU(2)$ isometry are 
\beq
\begin{split}
\vspace{3 mm}
&ds^2=G_{\mu\nu} dx^{\mu}dx^{\nu}+2G_{\mu i}dx^{\mu}L^i +g_ij L^iL^j\\
&B=B_{\mu\nu}dx^{\mu}dx^{\nu}+B_{\mu i}dx^{\mu}L^i+\frac{1}{2} b_{ij} L^i\wedge L^j
\end{split}
\eeq
where $L^{i}$ are $SU(2)$ left invariant 1-forms parametrised by Euler angles $\theta$, $\varphi$, $\psi$, along which the duality is performed and $\mu=0,1,...6$ are spectator coordinates. The dilaton is $\phi$ which can depend on any of the spectator coordinates.
A basis of vielbeins for such a solution can be written as
\beq
e^{A}= e^{A}_{\mu}dx^{\mu},~~~~ e^{a}= \kappa^{a}_{j}L^j+\lambda^{a}_{\mu}dx^{\mu}
\eeq
where $A=0,1,...,6$ and $a=1,2,3$.

The NS sector defines a sigma model with action given by
\beq
\mathcal{S}=\int d^2\sigma \bigg(Q_{\mu\nu}\del_+X^{\mu}\del_-X^{\nu} +Q_{\mu i} \del_+X^{\mu} L^i_- +Q_{i\mu} L^i_+\del_-X^{\nu}+E^{ij} L^i_+L^j_-\bigg)
\eeq
where
\beq
Q_{\mu\nu}=G_{\mu\nu}+B_{\mu\nu},~~~ Q_{\mu i}= G_{\mu i} + B_{\mu i},~~~Q_{i\mu}= G_{i\mu}+B_{i\mu},~~~ E_{ij}=g_{ij}+b_{ij}
\eeq
The T-dual sigma model can then be calculated by performing a Buscher procedure. This consists of gauging the $SU(2)$ isometry and adding a Lagrange multiplier term to impose flatness of the connection. After the multiplier term is integrated by parts the gauge fields can be solved for which gives the T-dual sigma model. At this point there are actually 3 more coordinates than required, namely $(\theta,\varphi,\psi,v_1,v_2,v_3)$, the redundancy must be eliminated by choosing a gauge. This may be parametrised by the matrix
\beq
D=\left(
\begin{array}{ccc}
 \cos \theta \cos \psi \cos \varphi-\sin \psi \sin \varphi & \cos
   \theta \sin \psi \cos \varphi+\cos \psi \sin \varphi & -\sin \theta
   \cos \varphi \\
 -\cos \theta \cos \psi \sin \varphi-\sin \psi \cos \varphi & \cos \psi
    \cos \varphi-\cos \theta \sin \psi \sin \varphi & \sin \theta \sin
   \varphi \\
 \sin \theta \cos \psi & \sin \theta \sin \psi & \cos \theta
\end{array}
\right)
\eeq
such that the dual coordinates given by
\beq
\tilde{v} = D^T v
\eeq
for $v=(v_1,v_2,v_3)$. The action of the T-dual sigma model can be written as
\beq
\tilde{\mathcal{S}}=\int d^2\sigma \bigg(Q_{\mu\nu}\del_+X^{\mu}\del_-X^{\nu}+(\del_+\tilde{v}_i+\del_+X^{\mu}Q_{\mu i})M_{ij}^{-1}(\del_-\tilde{v}_j-Q_{j\nu} \del_- X^{\nu})\bigg)
\eeq
and the Buscher rules defining the dual NS sector may be read off from
\beq
\begin{split}
&\tilde{Q}_{\mu\nu}= Q_{\mu\nu}-Q_{\mu i}M_{ij}^{-1}Q_{j \nu},~~~\tilde{E}_{ij}=M^{-1}_{ij}\\
& \tilde{Q}_{\mu i}= Q_{\mu j}M^{-1}_{ji},~~~ \tilde{Q}_{i \mu}=-M^{-1}_{ij} Q_{j\mu}
\end{split}
\eeq
where the dual dilaton and the matrix M are given by
\beq
e^{-2\tilde{\phi}}=\det M e^{-2\phi},~~~~~M_{ij}=E_{ij}+\epsilon_{ijk}\tilde{v}_k.
\eeq

The RR sector is calculated in a different way. It is possible to define both left and right moving dual vielbeins
\beq\label{eq: dualviel}
e\to \tilde{e}_+= -\kappa M^{-T}(dv+Q^{T}dX)+\lambda dX,~~~e\to \tilde{e}_-= \kappa M^{-1}(dv-QdX)+\lambda dX.
\eeq
But because they both define the same geometry these vielbeins must be equivalent up to a Lorentz transformations. Indeed they are and the transformation may be expressed as:
\beq
\Lambda=-\kappa^{-T}MM^{-T} \kappa^T
\eeq
this may then be used to define an action on spinors given by a matrix $\Omega$ by demanding that
\beq
\Omega^{-1}\Gamma^a\Omega=\Lambda^{a}_b \Gamma^b
\eeq
After some work, the details of which may be found in  \cite{Itsios:2013wd}, it can be shown that the solution to this constraint is given by
\beq
\Omega= \Gamma_{11} \frac{-\Gamma_{123}+\zeta_a\Gamma^a}{\sqrt{1+\zeta^2}}
\eeq
where
\beq\label{eq: zetas}
\zeta_a= \frac{\kappa^a_i(b^i+\tilde{v}^i)}{\det \kappa},~~~ b_{ij}=\epsilon_{ijk}b_{k}
\eeq
This gives the object needed to generate the dual RR sector. One constructs polyforms out of the RR fields of both type IIA and IIB SuGra
\beq
F_{IIA}=\sum_{n=0}^5F_{2n},~~~~~F_{IIB}=\sum_{n=0}^4F_{2n+1}
\eeq
where higher forms are related to lower forms via $F_n=(-1)^{\text{Int}[n/2]}\star F_{10-n}$.
These are then mapped to bispinors under the Clifford map: 
\beq
X=\frac{1}{p!}X_{\mu_1,...\mu_p}e^{\mu_1}\wedge...\wedge e^{\mu_p}\to\slashed{X}=\frac{1}{p!}X_{\mu_1,...\mu_p}\Gamma^{\mu_1,...\mu_p}
\eeq
where it is possible to make the identification
\beq
e^{\tilde{\phi}}\slashed{F}_{IIA}= e^{\phi} \slashed{F}_{IIB}\Omega^{-1}
\eeq
from which  the dual RR sector may be extracted.

One should realise that the method laid out in this section is not the only one that may be used to generate the NS and RR sectors of a non-abelian T-dual solution. An alternative is to use a consistent truncation to 7-d and match the original solutions there \cite{Itsios:2012dc}. Another interesting method is given by \cite{Gevorgyan:2013xka} where topological defects that generate the duality are constructed. This later method guarantees the Bianchi identities, but has not yet been shown to match the other methods in all cases. Finally this section has only discussed the dualising along $SU(2)$ isometries acting without isotropy, for more general isometries the interested reader is referred to \cite{Lozano:2011kb}.

\subsection{non-Abelian T-dual of wrapped D5 branes on $\Sigma^3$}
In this section a non-abelian T-duality is performed on the wrapped D5 brane solution of section \ref{Sec: II}. It acts along the $SU(2)$ isometry parametrised by $\omega^i$ and gauge fixing is imposed such that:
\beq
v_1=\theta_2=\phi_2=0.
\eeq

The dual NS sector is by,
\beq\label{eq: DualDil}
\begin{split}
e^{-2\Phi}&=\det M e^{-2\phi},\\
\det M&=\frac{1}{8} N_c^3 e^{6 g+3 \phi }+N_c e^{2 g+\phi }v_2^2 + N_c e^{2 g+\phi }v_3^2,
\end{split}
\eeq
for the dual dilaton $\Phi$,
\beq\label{eq: B2}
\begin{split}
B_2=&\frac{(w+1) N_c e^{2 g+\phi } \left(N_c^2 e^{4 g+2 \phi }+8 v_3^2\right) \left(\hat{\sigma}^1\wedge dv_2-dv_3\wedge
   \sigma^3\right)}{16 \sqrt{2} \det M}-\\
   &\frac{v_2 (w+1) N_c e^{2 g+\phi } \left(v_3 \hat{\sigma}^1\wedge dv_3-v_3 dv_2\wedge \sigma^3\right)}{2 \sqrt{2}
   \det M}+\frac{v_2 (w+1) N_c^3 e^{6 g+3 \phi } \hat{\sigma}^2\wedge  d\psi_2}{16 \sqrt{2}
   \det M}-\\
   &\frac{\left(v_2-v_3\right) \left(v_2+v_3\right) (w+1) N_c e^{2 g+\phi } \hat{\sigma}^1\wedge dv_2}{2 \sqrt{2} \det M}+\\
   &\frac{v_2 N_c e^{2 g+\phi }
   \left(v_2 dv_3\wedge  d\psi_2-v_3 dv_2\wedge  d\psi_2\right)}{\sqrt{2} \det M}-\\
   &\frac{(w+1)^2 N_c^3 e^{6
   g+3 \phi } \left(v_2 \hat{\sigma}^2\wedge \sigma^3+v_3 \hat{\sigma}^1\wedge \hat{\sigma}^2\right)}{32 \sqrt{2} \det M}
\end{split}
\eeq
for the NS two-form potential and
\beq
\begin{split}
d\hat{s}^2&= e^{\phi}\bigg[dx_{1,2}^2 +e^{2g}dr^2 +\frac{e^{2h}}{4}\left((\hat{\sigma}^1)^2+(\hat{\sigma}^2)^2+(\sigma^3)^2\right)\bigg]+\\
&\frac{1}{4\det M}\bigg[  v_2 v_3 (w+1) N_c^2 e^{4 g+2 \phi }\hat{\sigma}^1 \left(d\psi_2-\frac{1}{2} (w+1) \sigma^3\right)+\\ 
& ~~~~~~~~~~(w+1) N_c^2 e^{4 g+2 \phi }\hat{\sigma}^2 \left(v_3 dv_2-v_2 dv_3\right)+v_2^2 N_c^2 e^{4 g+2 \phi }
   \left(d\psi_2-\frac{1}{2} (w+1) \sigma^3\right)^2+\\
&~~~~~~~~~~\frac{1}{4}  v_3^2 (w+1)^2 N_c^2 e^{4 g+2 \phi } (\hat{\sigma}^1)^2+\frac{1}{4}  \left(v_2^2+v_3^2\right) (w+1)^2 N_c^2 e^{4 g+2 \phi }(\hat{\sigma}^2)^2+\\
&~~~~~~~~~~\left(N_c^2 e^{4 g+2 \phi }\left(dv_2^2+dv_3^2\right)+8 \left(v_2 dv_2+v_3 dv_3\right)^2\right)\bigg]
\end{split}
\eeq
for the dual metric. The new hatted 1-forms are simply a rotation in $\sigma^1$,$\sigma^2$:
\beq
\hat{\sigma}^1=\cos\psi_2\sigma^1-\sin\psi_2\sigma^2;~~~~\hat{\sigma}^2=\sin\psi_2\sigma^1+\cos\psi_2\sigma_2;
\eeq
that enables compact expressions.

The RR sector of the solution is rich including a quantised $F_0$ meaning that the solution is massive type-IIA. In order to express them compactly it is helpful to introduce the following basis of spectator vielbeins
\beq
\begin{array}{ll}
\vspace{3 mm}
e^{x^i}&=e^{\phi/2}dx^i, ~~~e^r= \sqrt{N_c}e^{\phi/2+2g}dr,~~~ e^{1,2}=\frac{\sqrt{N_c}e^{\phi/2+h}}{2}\hat{\sigma}^{1,2},~~~e^{3}=\frac{\sqrt{N_c}e^{\phi/2+h}}{2}\sigma^3\\
\end{array}
\eeq
with dual vielbeins given by
\beq
\begin{array}{ll}
\vspace{3 mm}
e^{\hat{1}}&=\frac{N_c^{3/2} e^{3 g+\frac{3 \phi }{2}}}{16 \det M} \bigg[4 \left(\hat{\sigma}^2 \left(v_2^2+v_3^2\right) (w+1)+2 \left(v_3 dv_2-v_2
   dv_3\right)\right)-\\
   \vspace{3 mm}
   &~~~~~~~~~~~~~~~~~~~~~~\sqrt{2} N_c e^{2 g+\phi } \left(\hat{\sigma}^1 v_3 (w+1)+2 v_2 \left(d\psi_2-\frac{1}{2} (w+1)
  \sigma^3\right)\right)\bigg],\\
e^{\hat{2}}&=\frac{\sqrt{N_c} e^{g+\frac{\phi }{2}}}{16 \det M} \bigg[-2 \sqrt{2} N_c^2 e^{4 g+2 \phi } \left(\hat{\sigma}^2 v_3 (w+1)+dv_2\right)-16 \sqrt{2} v_2 \left(v_2
  dv_2+v_3 dv_3\right)-\\
     \vspace{3 mm}
   &~~~~~~~~~~~~~~~~~~~~~~  4 v_3 N_c e^{2 g+\phi } \left(\hat{\sigma}^1 v_3 (w+1)+2 v_2 \left(d\psi_2-\frac{1}{2} (w+1)
  \sigma^3\right)\right)\bigg],\\
e^{\hat{3}}&=\frac{\sqrt{N_c} e^{g+\frac{\phi }{2}} }{16 \det M}\bigg[\sqrt{2} N_c^2 e^{4 g+2 \phi } \left(\hat{\sigma}^2 v_2 (w+1)-2 dv_3\right)-16 \sqrt{2} v_3 \left(v_2
  dv_2+v_3 dv_3\right)+\\
   &~~~~~~~~~~~~~~~~~~~~~~ 4 v_2 N_c e^{2 g+\phi } \left(\hat{\sigma}^1 v_3 (w+1)+2 v_2 \left(d\psi_2-\frac{1}{2} (w+1)
  \sigma^3\right)\right)\bigg],\\
  \end{array}
\eeq
which is a rotation of the rather complicated vielbeins generated by the duality procedure.
The fluxes are then:
\beq\label{eq: F}
\begin{array}{ll}
F_0&=\frac{N_c}{\sqrt{2}};\\
F_2&=-\frac{1}{4} e^{-2 (g+h)-\phi } \bigg[\sqrt{2} \left(-e^{3\hat{3}}+e^{2\hat{2}}+e^{1\hat{1}}\right) N_c (w-\gamma ) e^{3 g+h+\phi }-\\
&~~~8 e^{2 h} \left(v_3 e^{\hat{1}\hat{2}}-v_2
   e^{\hat{2}\hat{3}}\right)-2 e^{2 g} U \left(e^{23} v_2+e^{12} v_3\right)+\\
   &~~~2 e^{g+h} \left(\gamma ' \left(v_3 e^{r3}+v_2 e^{r1}\right)+2 (w-\gamma )
   \left(v_3 \left(e^{1\hat{2}}-e^{2\hat{1}}\right)-v_2 \left(e^{3\hat{2}}+e^{2\hat{3}}\right)\right)\right)\bigg];\\
F_4&=-\frac{(w-1)V e^{3g-3h}N_c}{8\sqrt{2}}e^{tx^1x^2r}+ \\
&\bigg[\frac{1}{4} V (w-1)( v_3e^{3\hat{3}}- v_2e^{e\hat{2}}) -\frac{1}{8} U e^{-2 h-\phi } \left(\sqrt{2} N_c e^{\hat{1}\hat{2}} e^{2 g+\phi }-4 v_2 e^{\hat{1}\hat{3}}\right)\bigg]\wedge e^{12}+\\
&\frac{1}{8} U e^{-2 h-\phi } \bigg[\sqrt{2} N_c e^{\hat{1}\hat{3}} e^{2 g+\phi }+4 v_2 e^{\hat{1}\hat{2}}+4 v_3 e^{\hat{2}\hat{3}}\bigg]\wedge e^{13}+\\
&\frac{\gamma'}{8} e^{-(g+h+\phi )} e^{r}\wedge\bigg[4 v_3 (e^{1\hat{1}\hat{3}}+e^{2\hat{2}\hat{3}})+4 v_2 e^{1\hat{1}\hat{2}}-e^{\hat{1}\hat{2}\hat{3}}-\\
&\sqrt{2} N_c (e^{1\hat{2}\hat{3}}-e^{2\hat{1}\hat{3}}-e^{3\hat{1}\hat{2}}) e^{2 g+\phi }\bigg]+\frac{1}{8} U e^{-2 h-\phi }e^{23\hat{3}} \left(\sqrt{2} e^{\hat{2}} N_c e^{2 g+\phi }-4 e^{\hat{1}} v_3\right)+\\
&e^{-(g+h+\phi )}(w-\gamma )\left(e^1 v_2+e^3 v_3\right)e^{\hat{1}\hat{2}\hat{3}};
\end{array}
\eeq
where the functions
\beq\label{eq: VU}
V=4+w^2-3 w \gamma +w-3 \gamma ,~~~~U=1+w^2-2 w \gamma;
\eeq
were introduced for convenience. Its interesting to see that $F_4$ has legs on the field theory directions. This tells us that, like the non-abelian T-dual of wrapped D5 branes on $S^2$ \cite{Itsios:2013wd}, the RR sector contains magnetic fluxes of D8, D6 and D4 branes, but here there is also an electric flux of D2 branes. 

It is simple to check that the RR fluxes automatically satisfy the Bianchi identities
\beq
d F_0= dF_2-F_0H=dF_4-H_3\wedge F_2=0.
\eeq
The flux equations of motion
\beq
d\star F_2+ H_3\wedge \star F_4=d\star F_4+ H_3\wedge F_4=0
\eeq
as well as Einsteins equations and the dilaton equation of motion all follow once eq \ref{eq: Gstr} is imposed. One may also confirm that NS flux obeys 
\beq
d\big(e^{-2\tilde{\phi}}\star H_3\big)=F_0\star F_2 +F_2\wedge \star F_4+\frac{1}{2} F_4\wedge F_4
\eeq 

\subsection{Supersymmetry}
In this section the issue of how much SUSY the non-abelian T-dual background preserves is addressed. There is a simple criterium which determines this, which is detailed in \cite{Sfetsos:2010uq,Itsios:2013wd}. One needs to consider the Kosmann derivative along each of the Killing vectors $k$ associated with the isometry on which one wishes to dualise. This acts on the Killing spinors of the initial solution $\epsilon$ and is given by
\be
\mathcal{L}_k \epsilon= k^{\mu}D_{\mu} \epsilon+\frac{1}{4} \nabla_{\mu} k_{\nu}\Gamma^{\mu\nu}\epsilon
\eeq
$D_{\mu}$ is the spinor covariant derivative while $\nabla_{\mu}$ is the ordinary covariant derivative of the geometry. The Kosmann derivative should vanish along the isometry of the dualisation. Each additional projective constraints that needed to be imposed to ensure this reduces the SUSY of the non-abelian T-dual by half. If no new constraints are required then all the SUSY of the original background is preserved.

The Killing vectors associated with the relevant $SU(2)$ isometry of the metric eq \ref{met} are,
\beq
\begin{split}
k^{(1)}&=-\cos\varphi_2\partial_{\theta_2}+\cot\theta_2\sin\varphi_2\partial_{\varphi_2}-\csc\theta_2 \sin\varphi_2\partial_{\psi_2}\\
k^{(2)}&=-\sin\varphi_2\partial_{\theta_2}-\cot\theta_2\cos\varphi_2\partial_{\varphi_2}+\csc\theta_2 \cos\varphi_2\partial_{\psi_2}\\
k^{(3)}&=- \partial_{\varphi_2}
\end{split}
\eeq
and it is possible to show that
\beq
\mathcal{L}_{k^{(1)}} \epsilon=\mathcal{L}_{k^{(2)}} \epsilon=\mathcal{L}_{k^{(3)}} \epsilon=0
\eeq
where $\epsilon$ only depends on $r$ (Appendix B of \cite{Canoura:2008at} provides further details). Thus one expects the non-abelian T-dual of wrapped D5 branes on $\Sigma^3$ to preserve $\mathcal{N}=1$ SUSY. 

A more explicit check that SUSY is preserved was provided by \cite{Barranco:2013fza}, which shows how non-abelian T-duality acts on the g-structure of the original geometry.  The original geometry supports a $G_2$-structure, which is characterised by parallel $\epsilon_1$ and $\epsilon_2$ in $\epsilon=(\epsilon_1,\epsilon_2)^T$. The SUSY conditions of the $G_2$ may be written in terms 2 real 7-d bispinors \cite{Fwitt}
\beq
\Psi_+=1-\star_7 \Phi_3,~~~~~\Psi_-=-\Phi_3+ Vol_7
\eeq
that obey the conditions \cite{Haack:2009jg}
\beq\label{eq:G2G2str}
\begin{split}
\vspace{3 mm}
&<\Psi_1,F>=0\\
\vspace{3 mm}
&(d-H_3\wedge)(e^{2A-\phi}\Psi_1)=0\\
&(d-H_3\wedge)(e^{3A-\phi}\Psi_2)+e^{3A}\star_7\lambda(F)=0.
\end{split}
\eeq
These are the conditions for $\mathcal{N}=1$ for a generic $G_2\times G_2$ structure manifold, where $e^{2A}$ is the warp factor of the Minkowski directions, $\lambda( X_p)=(-1)^{\frac{p(p-1)}{2}} X_p$ and $<A,B>$ is the Mukai pairing which selects the 7-d part of $X\wedge \lambda(Y)$. As the original solution is in type-IIB one should identify $\Psi_{1,2}=\Psi_{+,-}$, with the opposite identification made in type-IIA. The relevant observation of \cite{Barranco:2013fza} is that non-abelian T-duality acts on the bispinors of the geometry as
\beq
\begin{split}
\vspace{3 mm}
&\tilde{\slashed{\Psi}}_+=\slashed{\Psi}_-\Omega^{-1}\\
&\tilde{\slashed{\Psi}}_-=\slashed{\Psi}_+\Omega^{-1}
\end{split}
\eeq
it is possible to show (in Mathematica) that eq. \ref{eq:G2G2str} is satisfied with $\Psi_{1,2}=\tilde{\Psi}_{-,+}$ and $F$ and $H_3$ given by eq \ref{eq: F} and eq \ref{eq: B2} respectively, which shows that $\mathcal{N}=1$ SUSY is preserved.
The action of non-abelian T-dual on the 10-d MW Killing spinors is:
\beq\label{eq: MWspinors}
\tilde{\epsilon}_1=\epsilon_1,~~~~~\tilde{\epsilon}_2=\Omega.\epsilon_2
\eeq
which is a rotation. Since the $G_2$-structure of the original geometry requires that the spinors are parallel, the dual structure must 
be something more exotic. To identify the structure of the dual geometry it is  sufficient to calculate how the $\Omega$ matrix 
transforms the $G_2$-structure spinors. There exists a basis\footnote{This is a rotation of the basis of eq \ref{eq: orgviels} such that $e^1 \to \cos\alpha e^{1}-\sin\alpha e^{\hat{1}}$ and $e^{\hat{1}} \to \cos\alpha e^{\hat{1}}+\sin\alpha e^{1}$ with all other vielbeins unchanged.} such that the projections the original killing spinor obeys are given by
\beq\label{eq proj}
\Gamma_{r\hat{1}\hat{2}\hat{3}}\epsilon=\epsilon,~~~~ \Gamma_{1\hat{1}}\epsilon=\Gamma_{2\hat{2}}\epsilon=\Gamma_{3\hat{3}}\epsilon.
\eeq
One may decompose the 10-d geometry into a 3+7 split using an auxiliary 2-d space so that the gamma matrices are given by
\beq
\begin{split}
&\Gamma_{\mu}=\gamma_{\mu}\otimes \sigma_3 \otimes \mathbb{1}\\
&\Gamma_{a}= \mathbb{I}\otimes \sigma_1\otimes \gamma_a\\
&\Gamma^{(10)}=- \mathbb{I}\otimes \sigma_2\otimes \mathbb{I}\\
\end{split}
\eeq
where $\mu=0,1,3$ and $a=1,2,3,\hat{1},\hat{2},\hat{3}$. The killing spinor may be decomposed such that they have positive chirality as
\beq
\epsilon_{1,2}= \xi \otimes\left(
\begin{array}{c}
1\\
-i\\
\end{array}
\right)\otimes \chi
\eeq
where $\xi$ and $\chi$ are spinors in 3-d and 7-d respectively. In such conventions the 3 form associated to $G_2$ is then given by $\Phi_{abc}=-i \bar{\chi}\gamma_{abc}\chi$. It is possible to show that in this decomposition the T-dual Killing spinors are given by
\beq
\tilde{\epsilon}_1= \xi\otimes\left(
\begin{array}{c}
1\\
-i\\
\end{array}
\right)\otimes \tilde{\chi}_1,~~~~~\tilde{\epsilon}_2=\xi \otimes\left(
\begin{array}{c}
1\\
i\\
\end{array}
\right)\otimes \tilde{\chi}_2
\eeq
which have the correct chirality for type-IIA. Using the projections of eq \ref{eq proj} the 7-d spinors may be massaged into the form
\beq
\tilde{\chi}_1= \chi,~~~~\tilde{\chi}_2= \frac{\sin\alpha}{\sqrt{1+\zeta^2}} \chi + \sqrt{\frac{\cos^2\alpha +\zeta^2}{1+\zeta^2}} \chi^{\perp},
\eeq
where $\zeta^2 =\zeta_a\zeta^a$ ans $\zeta_a$ is given by eq \ref{eq: zetas}. As the notation suggests $\tilde{\chi}_2$ is a sum of parts which are parallel and orthogonal to $\tilde{\chi}_1$. The orthogonal complement to $\chi$ is 
\beq
\chi^{\perp} = i K \chi,~~~~ K= \frac{\cos\alpha \gamma_r+\zeta_1 \cos\alpha \gamma_{\hat{1}}+\zeta_2\gamma_{\hat{2}}+\zeta_3\gamma_{\hat{3}}+\zeta_1 \sin\alpha\gamma_\{hat{1}}{\sqrt{\cos^2\alpha +\zeta^2}}
\eeq
where K defines the 1-form associated with an $SU(3)$-structure in 7-d when contracted with the vielbeins of eq \ref{eq: dualviel}\footnote{The contraction should be performed once the vielbeins have been rotated to the canonical frame of footnote 4}. Specifically the structure is what should be called dynamical $SU(3)$, by analogy to dynamical $SU(2)$-structures in 6-d \cite{Heidenreich:2010ad}. This is because the coefficients of $\chi$ and $\chi^{\perp}$ are not constant through out the space, in fact $\sin\alpha\to 0$ as $r\to\infty$ so the structure becomes orthogonal $SU(3)$ in the UV, but through out the whole space the coefficients change. The details of this calculation and presentation of all the forms associated with $SU(3)$ shall be left for forthcoming work.
\subsection{Cycles and Charges}\label{sec: charges}
The T-dual geometry supports many fluxes and so contains several different branes. Since the internal space in 7-d, the possible quantised charges are given by
\beq
\begin{split}\label{eq: charges}
Q_{D6}&= \frac{1}{2\pi}\int_{\tilde{\Sigma}^2}F_2-F_0B_2,\\
Q_{D4}&=  \frac{1}{8\pi^3}\int_{\tilde{\Sigma}^4}F_4-B_2\wedge F_2+\frac{F_0}{2}B_2\wedge B_2,\\
Q_{D2}&=  \frac{1}{32\pi^5}\int_{\tilde{\Sigma}^6}F_6-B_2\wedge F_4+\frac{1}{2}B_2\wedge B_2\wedge F_2-\frac{F_0}{6}B_2\wedge B_2\wedge B_2.\\
\end{split}
\end{equation}
Sensible cycles over which to define these quantities are
\beq
\begin{split}
\tilde{\Sigma}^2&=\{\theta_1,\varphi_1| v_2=v_3= \psi_1=\psi_2=0\},\\
\tilde{\Sigma}^4&=\{\theta_1,\varphi_1,\psi_1,v_2| v_3=0,\psi_2=constant\},\\
\tilde{\Sigma}^6&=\{\theta_1,\varphi_1,\psi_1,v_2,v_3,\psi_2\}.\\
\end{split}
\eeq
Actually $\tilde{\Sigma}^2$ shrinks to zero in the IR, but as $F_2$ and $B_2$ vanish on this cycle this will not cause a singularity in the geometry as a non-zero Page charge must be pure gauge in origin. On these cycles eq. \ref{eq: charges} takes the simple form:
\beq
\begin{split}\label{eq: charges2}
Q_{D6}&= 0 ~\text{up to large gauge transformations} ,\\
Q_{D4}&=  \frac{1}{8\pi^3}\int_{\tilde{\Sigma}^4}\frac{N_c}{2\sqrt{2}}v_2\sin\theta_1 d\theta\wedge d\varphi_1\wedge d\psi_1\wedge dv_2,\\
Q_{D2}&= - \frac{1}{32\pi^5}\int_{\tilde{\Sigma}^6}\frac{N_c}{4}v_2\sin\theta_1 d\theta_1\wedge d\varphi_1\wedge dv_1\wedge dv_2\wedge \psi_2.\\
\end{split}
\eeq
to make further progress one needs to fix the periods of the dual coordinates $v_2,~v_3$. A rigorous prescription for doing this is absent form the literature, but it is at least reasonable to assume that they are compact. Here the periodicities shall be chosen such that
\beq
\int v_2 dv_2=\pi,~~~\int dv_3=\sqrt{2}\pi
\eeq
with these choices the D2 and D4 Page charges coincide with the Romans mass,
\beq\label{eq: commoncharge}
Q_{D4}=Q_{D2}=F_0= \frac{N_c}{\sqrt{2}}.
\eeq
since these objects contain explicit $B_2$ terms in their definitions they can experience quantised shifts under large gauge transformations $B_2\to B_2 +\Delta B_2$. For example
\beq\label{eq: B2 shift}
\Delta B_2 = \frac{n}{2}\sin\theta d\theta_1\wedge d\varphi_1,~~~ \Delta Q_{D6}=n\frac{N_c}{\sqrt{2}},~~~\Delta Q_{D4}=\Delta Q_{D2}=0.
\eeq

Finally there are 2 cycles on which the induced metric takes a particularly simple form. On $\tilde{\Sigma}^2$ the induced metric is given by
\beq\label{eq: Dual2cyc}
ds^2_{\tilde{\Sigma}^2}= N_c\frac{e^{2h+\phi}}{4}\big(d\theta_1^2+d\varphi_1^2\big)
\eeq
this cycle vanishes in the IR, blows up in the UV and $F_2$ and $B_2$ vanish on it. The second is the 3-sphere $S^3 = (\theta_1,\varphi_1,\psi_1)$ on which the metric is 
\beq\label{eq: Dual3cyc}
ds^2_{S^3}= N_c\frac{e^{2h+\phi}}{4}\big(d\theta_1^2+d\varphi_1^2+2\cos\theta_1 d\varphi_1d \psi_1+d \psi_1^2\big)
\eeq
which has the same asymptotic behaviour as the previous cycle.
\section{Probe Analysis and Comparison of the Gauge Theories}\label{sec: probes}
In this section some field theory observables shall be studied via a probe brane analysis. To begin, the results of \cite{Canoura:2008at} and \cite{Macpherson:2013gh} shall be reviewed to study the field theories dual to the wrapped D5-brane solution and its $G_2$-structure rotation respectively. A new proposal for the Chern-Simons level of $G_2$-structure rotated solution will be made before the non-abelian T-dual solution is considered.

The analysis of this section will rely on two important observations. They give a prescription for defining gauge couplings and Chern-Simons levels from probe branes.
\subsubsection*{\textit{ Gauge Coupling}}
A gauge coupling may be defined in terms of the DBI action of a probe Dp brane wrapping an n-cycle, where the embedding of this brane is $\xi=(t,x^1,x^2,\Sigma^n)$. In the following it shall be assumed that the induced metric can be expressed as
\beq \label{eq: emb1}
ds^2_{Dp}=e^{2A}dx^2_{1,2} + ds^2_{\Sigma^n}
\eeq
where its components $\hat{G}_{M,N}$ are decomposed as $M=(\mu,a)$ for $\mu=0,1,2$ and $a=1,...,p-2$. In addition the only non-zero part of the world volume gauge field $F$ and pull back of $B_2$ will be  
\beq \label{eq: emb2}
F_{\mu\nu},~~~~~\hat{B}_{ab}.
\eeq
and the dilaton shall be $\phi$. The DBI action of this probe Dp brane may be factorised into $\mathbb{R}^{1,2}$ and $\Sigma^{p-2}$ parts
\beq\label{eq: genDBI}
\begin{split}
S^{Dp}_{DBI}&=-T_{Dp}\int_{M_{p+1}} d^{p+1}\xi e^{-\phi}\sqrt{-\det \big(\hat{G}_{MN}+ \hat{B}_{NM} +2\pi \alpha'F_{MN}\big)}\\
&=-T_{Dp}\int_{\mathbb{R}^{1,2}}d^3x \sqrt{-\det (\hat{G}_{\mu\nu}+2\pi\alpha' F_{\mu\nu})}\int_{\Sigma^{p-2}}d^{(p-2)}\Sigma e^{-\phi} \sqrt{\det (\hat{G}_{ab}+\hat{B}_{ab})}.
\end{split}
\eeq
One may then expand the $\mathbb{R}^{1,2}$ determinant for small values of $\alpha'$, which leads to
\beq
\begin{split}
 \sqrt{-\det(\hat{G}+2\pi\alpha' F)}&= e^{3A}\sqrt{-\det(I+2\pi\alpha' \hat{G}^{-1}F)}\\
 &= e^{3A}\big[1-\frac{(2\pi\alpha')^2}{4} e^{-4A}tr(\eta^{-1} F\eta^{-1} F)+O(\alpha')^4\big]
 \end{split}
\eeq
where indices have been suppressed. $tr(\eta^{-1} F\eta^{-1} F)\equiv F_{\mu\nu}F^{\mu\nu}$ is the standard object appearing the in YM action
\beq
S_{YM}=\frac{1}{4g^2}\int d^3x F_{\mu\nu}F^{\mu\nu} 
\eeq
and so one may relate the $\alpha'~\!^2$ term in the expansion of eq \ref{eq: genDBI} to this coupling which leads to the identification
\beq
\frac{1}{g^2}= T_{Dp}\int_{\Sigma^{p-2}}d^{(p-2)}\Sigma e^{-\phi-A}\sqrt{\det(\hat{G}_{ab}+\hat{B}_{ab})}
\eeq
A second way one can define a gauge coupling is with a Euclidean Dp brane. The DBI action of such a brane will wrap a compact (p+1)-cycle $\Sigma^{p+1}$ in the internal space and is given by
\beq
S^{Dp}_{Euclid}= T_p\int_{\Sigma^p}d^{(p+1)}\S e^{-\phi}\sqrt{\det\big(\hat{G}_{ab}+\hat{B}_{ab}\big)}
\eeq
This can be identified with the action of an instanton
\beq
e^{-S_{inst}}=e^{-\frac{8\pi^2}{g^2}}
\eeq
Thus for a Euclidean Dp brane can give a gauge coupling  
\beq
\frac{8\pi^2}{g^2} =T_p\int_{\Sigma^p}d^{(p+1)}\S e^{-\phi}\sqrt{\det\big(\hat{G}_{ab}+\hat{B}_{ab}\big)}
\eeq
In 4-d one would also include the WZ term to define a $\Theta$ angle. However the ideas behind this do not necessarily extend to 3-d so this term will be ignored here.
\subsubsection*{\textit{Chern-Simons level}}
A Chern-Simons level can be be extracted from the WZ action of a Dp brane with embedding $\xi=(t,x^1,x^2,\Sigma^n)$ in a similar fashion. The exact prescription is dependent on what conventions are being used. For instance in conventions such that the RR ployform may be expressed as
\beq
F_{poly}=dC-H_3\wedge C + F_0 e^{B_2}
\eeq
where C is the sum over the potentials of type-IIA or type-IIB (Additionally $F_0$ should be taken to be zero in this case). Define also the Page charge of D(10-p) brane on a $(p-2)$-cycle $\Sigma^{(p-2)}$ to be
\beq
Q^{D(10-p)}= \frac{s_{(10-p)}}{2\kappa_{10}^2 T_{(10-p)}}\int_{\Sigma^{(p-2)}}\big(F_{poly}\wedge e^{-B}-F_0\big),
\eeq
 where $-F_0$ ensures that the D8 brane Page charge, which is really a mass is not included. The orientation of the cycle is parametrised by $s_{(10-p)}=\pm1$. Finally the WZ action of a Dp-brane shall be given by
\beq\label{eq: WZ}
S^{Dp}_{WZ}=s'_pT_p\int_{M_{p+1}}C\wedge e^{-\hat{B}_2- 2\pi\alpha' F}
\eeq 
where the action is allowed to come with an overall positive or negative sign, i,e. $s'_p=\pm1$.
A Chern-Simons term in a gauge theory is given by the action
\beq
S_{CS}=-\frac{k}{4\pi}\int d^3x\mathcal{L}_{CS} =-\frac{k}{4\pi}\int d^3x tr(dA\wedge A + \frac{2}{3}A\wedge A \wedge A)
\eeq
where $A$ is a gauge field with field strength $F= dA + A\wedge A$ and so  $d\mathcal{L}_{CS}=F\wedge F$. The order $F\wedge F$ term in \ref{eq: WZ} may be manipulated by adding an exact to give a Chern-Simons term,
\beq
\begin{split}
d\big[C\wedge e^{-B_2}\wedge \mathcal{L}_{CS}\big]\pm C\wedge e^{-B_2}\wedge F\wedge F&= \big(dC-H_3\wedge C\big)\wedge e^{-B_2}\wedge \mathcal{L}_{CS}\\
&=\big(F_{poly}\wedge e^{-B_2}-F_0\big)\wedge \mathcal{L}_{CS}.
\end{split}
\eeq
where the $+/-$ sign is for IIA/IIB. The Chern-Simons term is then given by
\beq
\begin{split}
S_{CS}&=s'_p\frac{(2\pi \alpha')^2}{2}T_p\int_{\Sigma^{(p-2)}}\big(F_{poly}\wedge e^{-B_2}-F_0\big)\int_{\mathbb{R}^{1,2}} \mathcal{L}_{CS}\\
&= -\frac{Q^{D(10-p)}}{4\pi}\int_{\mathbb{R}^{1,2}}\mathcal{L}_{CS}
\end{split}
\eeq
where $s'_p s_{10-p}=-1$ and  $2\kappa_{10}^2 T_pT_{10-p}= (2\pi)^{-3}$ have been used. This shows quite generally that the WZ action of a Dp brane contains a Chern-Simons coupling of level $Q_{10-p}$. Actually this is not the whole story as the true Chern-Simons level can experience an additional shift when all the $p$-dimensional KK modes are integrated out. Extra care must be taken when different conventions are used, indeed this is the case in the $G_2$-structure rotated solution, where further details will be given.
\subsection{Wrapped D5 Branes and $\mathcal{N}=1$ SYM with Gauge Group $SU(N_c)_{\frac{N_c}{2}}$}
The solution of wrapped D5 Branes on $\Sigma^3$ is dual in the IR to $\mathcal{N}=1$ SYM in 3 dimensions. It contains $N_c$ color branes as can be seen from the flux quantisation condition
\beq
-\frac{1}{4\pi^2}\int_{\tilde{S}^3} F_3 =N_c 
\eeq
so the gauge group is $SU(N_c)$. The geometry only gives a good holographic description of a field theory in 3-d in the IR where $r\sim0$. This is because $\Sigma^3$ vanishes in the IR and the QFT living on the wrapped D5 branes is effectively 3 dimensional there, however in the UV the cycle blows up and the world volume is explicitly 6 dimensional.

A suitable definition of the gauge coupling is given by a probe D5-brane extended along Minkowski and wrapping $\Sigma^3$. Once a gauge field $F$ with legs in the Minkowski directions is turned on, the action of such a brane is given by
\beq
S_{probe}= T_5\int_{\mathbb{R}^{1,2}\times \Sigma^3}d^3x d\Sigma^3e^{-\phi}\sqrt{-\det(G_{ind}+2\pi\alpha'F)},
\eeq
there is no WZ contribution as $F_3=0$ on $\Sigma^3$. At this stage it will be instructive to reintroduce $g_s$ and $\alpha'$ so that the induced metric is given by
\beq
ds^2_{ind}=e^{\phi}\bigg[dx_{1,2}^2 +\frac{\alpha'g_s N_c}{4}(e^{2h}+\frac{e^{2g}}{4}(w-1)^2)(\sigma^i)^2\bigg]
\eeq
this and the fact that $(2\pi)^2\alpha'\!~^3 g_sT_5=1$ gives the following $\alpha'$ expansion of the DBI action,
\beq
S_{probe}=-\frac{\sqrt{g_s N_c} N_c}{16\pi^3 \alpha'\!~^{3/2}}e^{2\phi}\big(e^{2h}+\frac{e^{2h}}{4}(w-1)^2\big)^{3/2}\int d^3x\big[1-2\pi^2e^{-2\phi}F_{\mu\nu}F^{\mu\nu}\alpha'\!~^2+...\big]
\eeq
where indices are contracted with the Minkowski metric. One can then identify $F^2$ term with the Yang-Mills action and make the identification
\beq\label{eq: coupling 1}
\frac{2\pi}{g^2N_c}=\sqrt{g_s N_c \alpha'} \big(e^{2h}+\frac{e^{2g}}{4}(w-1)^2\big)^{3/2}.
\eeq
which gives a coupling of mass dimension 1, as it should have in 3-d. The RHS of eq. \ref{eq: coupling 1} blows up in the UV and vanishes in the UV, which is consistent with the asymptotic freedom and confinement on expects of SYM in 3-d. The second of these is further supported by a Wilson loop calculation as in \cite{Canoura:2008at,Macpherson:2012zy}, which gives an area law with string tension $\sigma=\frac{1}{2\pi \alpha'}e^{\phi_0}$.

One can also calculate the Chern-Simons level from a probe brane. Consider a D5-brane extended along Minkowski and wrapping  $S^3$, the WZ action of such a brane is 
\beq
S_{WZ}=T_5\int_{\mathbb{R}^{1,2}\times S^3}\big(C_6+(2\pi\alpha')^2C_2\wedge F\wedge F),
\eeq
where $F$ is once more a world volume gauge field with legs in the field theory directions. Integrating the second term in this action by parts gives the Chern-Simons action \cite{Maldacena:2001pb}
\beq
-\frac{1}{16\pi^3}\int_{S^3} F_3 \int d^3xtr\big(dA+\frac{2}{3}A\wedge A\wedge A\big) = -\frac{k_6}{4\pi}\int d^3x\mathcal{L}_{CS}.
\eeq
where $k_6=N_c$. There is no $g_s$ or $\alpha'$ factors because they cancel with those in $F_3$ once they are reimposed. The $k_6$ here is to distinguish this object from the true CS level $k$ which gets an extra contribution when one integrates out all the 6-d Kaluza Klein modes\footnote{Specifically this it is integrating the massive fermions that generates the shift \cite{Maldacena:2001pb}} . The Chern-Simons level is then
\beq
k=k_6-\frac{N_c}{2}= \frac{N_c}{2}
\eeq
which is half integer as one expects due to the parity anomaly in 3-d. 

The wrapped D5-brane solution has two distinct UV solutions characterised by an asymptotically linear and constant dilaton, i.e. eq \ref{eq: UV1} and eq \ref{eq: UV2}. The field theoretic interpretation for this is that the constant dilaton solutions have an irrelevant operator insertion in their Lagrangian. 
\subsection{The $G_2$-structure Rotation and a 2-node Quiver Chern-Simons Theory}\label{sec: g2rotprobes}
In \cite{Macpherson:2013gh} much of the gauge theory analysis of the $G_2$-structure rotated solution was performed. It was concluded that the gauge group was that of a 2 node quiver of the form $ SU(r_l)\times SU(r_s)$, where $r_l>r_s$, by analogy with the baryonic branch of Klebanov-Strassler. The objects that must give rise to the ranks of this product group are the Page charges of the D2 and NS5 branes. As explained at length in \cite{Benini:2007gx}, under a Seiberg duality the ranks of the gauge groups transform as $r_l'= r_s$ and $r_s'=2r_s-r_l$. It is possible to see this manifestly in the supergravity solution if one defines
\beq\label{eq: defnofranks}
Q_{NS5}=r_l-r_s,~~~~~Q_{D2}=r_s.
\eeq
The Page charges on the LHS of these equalities transform under the large gauge transformation of eq \ref{eq: gaugetrans} in precisely the same way as the ranks on the RHS do under a Seiberg duality. Thus large gauge transformations are equivalent to Seiberg duality. This along with the fact that there is a running integral of $C_3$ at infinity \cite{Macpherson:2013gh} is very suggestive of a duality cascade, once more by analogy with Klebanov-Strassler. It is reasonable then to propose that in the UV, where $Q_{D2}=M_c$, the gauge group is $ SU(N_c+M_c)\times SU(M_c)$ and this then cascades down in ranks as one flow towards the IR terminating at $SU(N_c)$ as Klebanov-Strassler does. 

It is possible to define two couplings for this quiver\footnote{In \cite{Macpherson:2013gh} a 3rd coupling is also proposed in terms of a D2 instanton, however this is probably not a good definition because the the WZ term is not quantised}, $g_1$ and $g_2$ in the same spirit as in the previous section. A probe D4 brane with $(t,x^1,x^2,\hat{\Sigma}^2)$, with $\hat{\Sigma}^2$ as in eq \ref{eq: sigmahat2} defines a coupling
\beq\label{eq: couplinghat1}
\frac{4 \pi^2}{g_1^2N_c}=\sqrt{\alpha'}e^{\phi_{\infty}-\phi^{0}}\sqrt{H}\big(e^{2h}+\frac{1}{4}e^{2g}(w-1)^2\big).
\eeq
while a probe D2 brane extended in Minkowski can be used to define the coupling
\beq\label{eq: couplinghat2}
\frac{1}{g_2^2}=\frac{\sqrt{\alpha'}}{g_s}e^{\phi_{\infty}-\phi^{0}}
\eeq
where both of these couplings have mass dimension 1 as they should. The LHS of eq \ref{eq: couplinghat1} vanishes at $r\sim 0$ and becomes constant as $r\to\infty$. This indicates that the coupling $g_1$ is consistent with confinement in the IR and dilation invariance in the UV. On the other hand the LHS of eq \ref{eq: couplinghat2} interpolates between a smaller and larger constant between the IR and the UV respectively. In \cite{Macpherson:2013gh} the difference in the IR behaviour of $g_2$ was interpreted as a signal of a confining Chern-Simons term dominating the gauge theory dynamics there. This can be understood because in a YM-CS like theory the level $k$ induces an effective mass for the gauge field,  $g^2_{YM}|k|$, which causes the Yang-Mills coupling to freeze at a constant value in the IR. Further evidence of confinement is given by Wilson loop calculations which obey an area law with string tension $\sigma=(c \sqrt{1-e^{2(\phi_{\infty}-\phi_0)}})^{-1}$. However a proposal for the Chern-Simons term which is claimed to be dominating the dynamics in the IR has been absent from the literature until now, the expression is provided below.

Indeed consider a probe D8 brane with embedding $(t,x^1,x^2,\hat{\Sigma}^6)$ on which a world volume gauge field is turned on with support in the Minkowski direction, the order $F\wedge F$ term of the WZ action of such a brane is
\beq\label{eq: LWZD8}
S_{CS}=-\frac{(2\pi\alpha')^2T_8}{2}\int_{\mathbb{R}^{1,2}\times \Sigma^6}\big(C_5+ B_2\wedge C_3)\wedge F\wedge F
\eeq
the integrand can be manipulated by adding an exact 
\beq
\big(C_5+ B_2\wedge C_3)\wedge F\wedge F+d\big[(C_5+ B_2\wedge C_3)\wedge \mathcal{L}_{CS}\big]
=\big(F_6+H_3\wedge C_3)\wedge \mathcal{L}_{CS}
\eeq
where $F_6=dC_5+ B_2\wedge F_4$, $F_4=dC_3$ and $F\wedge F=d\mathcal{L}_{CS}$ have been used. Plugging this back into eq. \ref{eq: LWZD8} and taking note of the definition of the D2 Page charge in eq. \ref{eq: NS5andD2charges} gives
\beq\label{eq: lcs}
S_{CS}=-\frac{Q_{D2}}{4\pi}\int_{\mathbb{R}^{1,2}}\mathcal{L}_{CS}
\eeq
thus if one takes into account the definitions of the ranks in eq \ref{eq: defnofranks}, a Chern-Simons level can be defined which is equal to the rank of smaller group
\beq
\hat{k}=r_s.
\eeq
Of course it is possible that the level will experience a shift when one integrates out the 8-d KK modes, therefore this result should be viewed as correct up to the possible effect of this subtlety. Clearly $\hat{k}$ is not a fixed number from the perspective of supergravity, it shifts under large gauge transformations of $C_3$, however it is always quantised as a Chern-Simons level must be. In eq \ref{eq: lcs} only the positive orientation of $\hat{\Sigma}^6$ is considered, indeed it is possible to define another with the negative orientation, ie $k_1=-k_2=k$. This is what happens in the ABJM \cite{Aharony:2008ug} where the $AdS_4\times S^7/\mathbb{Z}_k$  geometry in M-theory is dual to the 2 node quiver $SU(N)_{-k}\times SU(N)_{k}$. This suggests that the quiver of the rotated solution could be $SU(r_l)_{-r_s}\times SU(r_s)_{r_s}$ by analogy with ABJM. If this is correct the effect of the Seiberg-like duality of eq \ref{eq: defnofranks} on the field theory is such that
\beq
G=SU(N_c+M_c)_{-\hat{k}}\times SU(M_c)_{\hat{k}}~~, ~~~~\hat{k}=M_c
\eeq
becomes
\beq
G'=SU(M_c-N_c)_{-\hat{k}'}\times SU(M_c)_{\hat{k}'}~~, ~~~~\hat{k}'=M_c-N_c
\eeq
and so clearly any cascade in the ranks of the groups must be associated with a corresponding cascade in the Chern-Simons levels.

The $G_2$-structure rotation acts on the SuGra solution that is dual to a QFT with an irrelevant operator that dominates the UV. The rotation induces additional warping on the metric by the function $H$ which makes the new metric asymptotically $AdS_4$, this means that the rotated solution no longer contains this operator. These warp factors also pre-multiply the internal space $ds^2_7$ and ensure that in the UV this remains finite. The field theory lives on the world volume of D2 and fractional D2 branes which do not unwrap like the D5 branes of the original solution, so the rotated solution gives a good holographic description of a 3-d gauge theory throughout the whole space. In this sense the rotation procedure can be seen as providing a UV completion to the original QFT dual to the wrapped D5 solution with asymptotically constant dilaton.
\subsection{The non-Abelian T-dual: Probe Analysis}
The geometry of the non-abelian T-dual of the wrapped D5 solution supports all possible fluxes. This fact and comparison to the rotated solution suggest that the field theory is a type of quiver. As discussed in section \ref{sec: charges}, it is possible to define several quantised charges once the periods of the dual coordinates $v_2$ and $v_3$ are fixed. Note however that the charges defined in eq \ref{eq: commoncharge} have a common $\sqrt{2}$ factor. This is an artefact of the conventions used in the dualisation procedure, it has no deep meaning and so it makes sense to make the redefinition
\beq
\tilde{N}_c=\frac{N_c}{\sqrt{2}}
\eeq
where $\tilde{N}_c$ should now be thought of as integer valued. The Page charges supported by the dual geometry are then 
\beq\label{eq: commoncharge2}
Q_{D6}=0 ~mod~\tilde{N}_c,~~~~~ Q_{D4}=Q_{D2}=\tilde{N}_c.
\eeq
This is enough charges to potentially define a 3 node quiver, but the identification of this quiver will not be pursued here, instead this section will focus only on probing the dynamics of the dual gauge theory. These probe calculation in the T-dual solution will be more complicated than the previous examples, for that reason the units as well as multiplicative constants in the couplings will be suppressed.

Like the rotated solution it is possible to define a coupling via a D2 brane parallel to the field theory coordinates. The dilaton, which is expressed in eq \ref{eq: DualDil}, depends on the dual coordinates that shall be set to constant values on the world volume of the brane. The simplest choice is that $v_2=v_3=0$ for which the coupling is given by
\beq
\frac{1}{\tilde{g}^2_1}\sim e^{3g}=\left\{\begin{array}{l l}\big(1 ~,~~~g_0^{3/2}\big)&~~~~~~~~~~~~r\sim0\\[3 mm]\big(1~, ~~~c^{3/2}e^{2r}\big)&~~~~~~~~~~~~r\to\infty\\ \end{array}\right.
\eeq
where the brackets correspond to asymptotically $\big($Linear, Constant$\big)$ dilaton solutions. The coupling is constant for the linear dilaton solution however this is clearly not a sign of conformal invariance as the non-compact dual metric is not $AdS_4$ and the dilaton is not constant. The coupling for solutions with constant dilaton in the UV are more interesting, asymptotically it is free and increases as one flows towards the IR finally freezing at a constant at $r=0$.

Another way to define a gauge coupling is a probe D4 brane with embedding $(t,x^1,x^2,\tilde{\Sigma}^2)$, where the induced metric is given by eq \ref{eq: Dual2cyc}. $B_2$ as defined in eq \ref{eq: B2} vanishes on this cycle, however non-vanishing contributions can be induced by large gauge transformations as in eq \ref{eq: B2 shift}. Generically the $F^2$ contribution to the DBI action gives a coupling of the form
\beq\label{eq: tildeg2}
\begin{split}
\frac{1}{\tilde{g}^2_2}= T_{D4}\int_{\tilde{\S}^2}e^{-3\phi/2}\sqrt{\det\big(\hat{G}_{ab}+\hat{B}_{ab}\big)}&\sim \frac{e^{3g}}{\pi}\int^{\pi}_0 d\theta_1\sqrt{e^{4h+2\phi}\tilde{N}_c^2+2 n^2 \sin^2\theta_1}\\
&= \frac{2N_c}{\pi}e^{3g+2h+\phi} E\big(\frac{-n^2}{e^{4h+2\phi} \tilde{N}_c^2}\big)
\end{split}
\eeq
where $\hat{B}= n/2 \sin\theta_1 d\theta_1\wedge d\varphi_1$, to include the effect of large gauge transformations. The function $E$ is a complete elliptic integral, which as a statement is not very illuminating. When $n=0$ there is no gauge transformation and $E(0)=\pi/2$ so the coupling is simply
\beq
\frac{1}{\tilde{g}^2_2}\sim e^{3g+2h+\phi}=\left\{\begin{array}{l l}\big(e^{\phi_0}r^2 ~,~~~g_0^{5/2}e^{\phi_0}r^2\big)&~~~~~~~~~~~~r\sim0\\[3 mm]\big(2 e^{\phi_{\infty}}r~, ~~~\frac{3c^{5/2}e^{\phi_{\infty}}}{4}e^{10r/3}\big)&~~~~~~~~~~~~r\to\infty\\ \end{array}\right.
\eeq
where the brackets correspond to asymptotically (Linear, Constant) dilaton solutions. This coupling is consistent with confinement in the IR and asymptotic freedom in the UV, in fact it is much like the coupling of the original background in eq \ref{eq: coupling 1}. For non-zero values of $n$ the UV behaviour is unchanged because $e^{4h+2\phi}$ becomes large and the elliptic integral is well approximated by
\beq
E(\frac{-n^2}{e^{4h+2\phi} \tilde{N}_c^2})\sim \frac{\pi}{2}+\frac{e^{-4h-2\phi}n^2}{4\tilde{N}_c^2},
\eeq
where the second term vanishes as $r\to\infty$. The IR behaviour changes quite dramatically under large gauge transformations because for $r\sim0$, $\sqrt{e^{4h+2\phi}N_c^2+2 n^2 \sin^2\theta_1}\sim\sqrt{2}N_c n\sin\theta_1$ and so the coupling tends to
\beq
\frac{1}{\tilde{g}_2^2}\bigg|_{r\sim0} \sim \frac{2\sqrt{2}|n|g_0^{3/2}}{\pi}
\eeq 
so the effect of the large gauge transformation is to freeze the coupling in the IR making $n=0$ a special case.

A third coupling may be defined in terms of Euclidean $D2$ on $\tilde{S}^3$. $B_2$ vanishes on this cycle up to the same large gauge transformations as before and the induced metric is given by eq \ref{eq: Dual3cyc} which leads to the coupling
\beq\label{eq: tildeg3}
\frac{1}{\tilde{g}^2_3}\sim e^{3g+h+\phi} \sqrt{2n^2+ e^{4h+2\phi}\tilde{N}_c^2}\sim\left\{\begin{array}{l l l}\big( e^{2\phi_0}r^3~,~~~e^{2\phi_0}(g_0 r)^3\big)&~~~~~~~~~~~~r\sim0,~n=0\\[3 mm]\big(~e^{2\phi_{\infty}}r^{3/2}~, ~~~c^3 e^{4r}\big)&~~~~~~~~~~~~r\to\infty\\[3 mm]
~~~g_0^2|n| \sqrt{r}&~~~~~~~~~~~~r\sim0,~n\neq0
\end{array}\right.
\eeq
this coupling is consistent with a strong coupling in the UV and asymptotic freedom in the IR. The effect of the large gauge transformation is less pronounced than it was for $\tilde{g}_2$ the behaviour in the IR is modified such that the power law changes, but the RHS of eq \ref{eq: tildeg3} still tends to zero in the IR.

The confining behaviour of the $\tilde{g}_3$ coupling should come as no surprise, the field theory and holographic directions are the same in both the original and dual geometries and so the conclusion of confinement from the Wilson loop studies of \cite{Canoura:2008at,Macpherson:2012zy} transfer to this solution also. That all the coupling exhibit asymptotic freedom is tied up with the fact that the bad UV behaviour of the original geometry, fractional branes unwrapping in the UV, is not being fixed by the T-duality, the irrelevant operator of the asymptotically constant dilaton solution will also still be present. As the original wrapped D5 brane solution is dual to a gauge theory with a Chern-Simons term it is reasonable to expect that the non-abelian T-dual geometry will be dual to a theory that also contains this type of term. That the couplings $\tilde{g}_1$ and (after a large gauge transformation) $\tilde{g}_2$ freeze out in the IR is certainly suggestive of a Chern-Simons term (or terms) that dominate the physics there.

At the beginning of this section it was shown that it is possible to define a Chern-Simons level for each Page charge in the geometry. For the non-abelian T-dual solution this gives 3 possible definitions
\begin{itemize}
\item Probe D8 brane on $(t,x^1,x^2,\tilde{\Sigma}_6)$ gives $k_1=Q_{D2}=\tilde{N}_c$\\[-4 mm]
\item Probe D6 brane on $(t,x^1,x^2,\tilde{\Sigma}_4)$ gives $k_2=Q_{D4}=\tilde{N}_c$\\[-4 mm]
\item Probe D4 brane on $(t,x^1,x^2,\tilde{\Sigma}_2)$ gives $k_3=Q_{D6}=n\tilde{N}_c$
\end{itemize}
up to possible shifts from integrating out all the massive KK modes. The $n$ in the definition of $k_3$ comes from the large gauge transformation $B_2=n/2 \sin\theta_1 d\theta_1 \wedge d\varphi_1$. The Chern-Simons level defined in the wrapped D5-brane solution is calculated on the 3-cycle $S^3=(\sigma^1,\sigma^2,\sigma^3)$. This cycle is orthogonal to the directions on which the dualisation is performed and so must be mapped to a 4-cycle and 6-cycle. This accounts for $k_1$ and $k_2$ and suggests that the D8 and D6 branes may be probing the same gauge theory object. $k_3$ is unambiguously distinct, it is zero when $B_2$ is defined as in eq \ref{eq: B2} but is shifted by large gauge transformation which is analogous to the Chern-Simons level of the $G_2$-structure rotated solution. It is interesting to see that when $k_3=0$ the couplings $\tilde{g}_2$ and $\tilde{g}_3$ behave quite differently than when it is not. Most pronounced is the effect on $g_2$ that exhibits typical confining behaviour when $n=0$ but freezes in the IR becoming constant otherwise. This can be interpreted as a clear example of the effect a non-zero Chern-Simons term can have on a Yang-Mills coupling, see the discussion below eq \ref{eq: couplinghat2}.

\section{Concluding Remarks}\label{sec: conclusion}
In this work the results of applying 2 solution generating techniques to the wrapped D5 solution of Maldacena and Nastase \cite{Maldacena:2001pb} (and its deformation \cite{Canoura:2008at}) were studied, with the aim of better understanding the dual gauge theories that are generated. 

The first technique, $G_2$-structure rotation  \cite{Gaillard:2010gy}, which is equivalent to  U-duality, has an action on the field theories that is already quite well understood, partially due to explicit calculation \cite{Gaillard:2010gy,Macpherson:2013gh,Macpherson:2012zy,Macpherson:2013dta} and partially by analogy to its 6-d $SU(3)$-structure equivalent \cite{Gaillard:2010qg,Elander:2011mh,Bennett:2011va,Conde:2011aa,Maldacena:2009mw,Caceres:2011zn}. The rotation acts on the wrapped D5 solution with asymptotically constant dilaton which is dual to a $\mathcal{N}=1$ SYM-CS in 3-d with an irrelevant operator insertion. After the rotation this operator is removed and the metric is asymptotically $AdS_4\times Y$, where $Y$ has finite volume. It was shown in \cite{Gaillard:2010gy,Macpherson:2013gh} that the rotated geometry is dual to a 2-node quiver that very likely exhibits a duality cascade like the baryonic branch of Klebanov-Strassler \cite{Klebanov:2000hb}, due to similarities between the two solutions. One way in which the solutions differ is that the $G_2$-structure rotated, being a holographic description of a 3-d QFT, can contain a Chern-Simons term. Evidence for this was given in \cite{Macpherson:2013gh} where, through a probe brane calculation, a YM  coupling was shown to freeze in the IR. This was interpreted as a signal of a Chern-Simons term that was dominating the IR, but no proposal for the level of this theory was given. This is resolved in section \ref{sec: g2rotprobes} where it is shown that a probe D8 brane wrapping the whole compact part of the rotated $G_2$-manifold, gives rise to a Chern-Simons level which is equal to the D2 Page charge. Thus the putative duality cascade must be accompanied by a cascade in the Chern-Simons level, which is very interesting and deserves further study. This will be left for future study as the main purpose of introducing the rotated solution was to aid, by comparison, the understanding of the main focus of this work.

A non-abelian T-duality \cite{delaossa:1992vc,Sfetsos:2010uq,Lozano:2011kb,Itsios:2012dc} was performed on the $SU(2)$ isometry parametrised by the left invariant 1-forms $\omega^i$ of the deformed Maldacena-Nastase solution. The result of this is a rather complicated solution in massive type-IIA with all possible RR forms turned on. As the duality does not change the directions orthogonal to the isometry, it does not improve the asymptotic behaviour of the field theory directions and holographic coordinate as the rotation does, this of course was to be expected. It was possible to explicitly show that under the T-duality the $G_2$-structure of the original solution is mapped to a dynamic $SU(3)$-structure in 7-d \cite{Gauntlett:2002sc,Martelli:2003ki}. This is the analogue of result of \cite{Barranco:2013fza} where it was shown that the 6-d $SU(3)$-structure of Klebanov-Witten \cite{Klebanov:1998hh} is mapped to a static $SU(2)$-structure for the T-dual solution \cite{Itsios:2012zv,Itsios:2013wd}. Indeed the structure of the dual geometry considered here becomes static $SU(3)$ in a limit in which, like Klebanov-Witten, there is no rotation in the projections of the original background (see Appendix A of \cite{Canoura:2008at} for details of the projections)

A rigorous prescription for fixing the periodicities of the dual coordinates is lacking. The view taken in this work was that the coordinates were at least likely to be compact. If this were not the case it would only be possible to define a D6 brane Page charge and this seems strange given the rich variety of fluxes. Periods were chosen for the dual coordinates such that Page charges for D2 and D4 branes could also be defined and such that these charges were equal. It is important to realise however that it should be possible to fix the periods of the dual coordinates by some requirement on the global properties of the dual manifold and that such a prescription may not match the choice made here. At any rate, it is unlikely that the specific choice would drastically change the salient features of the manifold so a probe analysis of the geometry was performed with periodicity thus fixed to gain some insight into the possible dual QFT.

That it is possible to define 3 Page charges suggests, by analogy with the rotated solution, that the dual gauge theory may be a 3 node quiver. However, unlike the rotated solution, it is possible in this case to define as many Chern-Simons terms as there are charges. The most interesting of these is the Chern-Simons term with level that coincided with the D6 Page charge $k_3$. This experiences shifts under large gauge transformations of $B_2$ in much the same way as is true for the level in the rotated solution (albeit with that shift mediated by large gauge transformations of $C_3$). A new feature in the T-dual solution is that the large gauge transformation actually changes the IR behaviour of two of the couplings that can be defined, this is most pronounced for the coupling $\tilde{g}_2$. When $k_3=0$ the coupling exhibits typical confining behaviour, it tends to infinity as when flows towards the IR. However when $k_3\neq0$ the coupling freezes in the IR, a sign that a confining Chern-Simons term is now dominating the dynamics this coupling is sensitive to. This constitutes a very clean example of such behaviour, and it is nice to see a familiar dynamical effect in this complicated SuGra solution. It is probable that this solution also experiences some shift in the ranks of gauge groups of the QFT and that it can perhaps be identified with a Seiberg/level-rank like duality, this will be left for future work. 

The general outlook for non-abelian T-duality as a SuGra solution generating technique with applications to holography seems good. This work represents a concrete example where interesting dynamics are generated by the duality. Given the wide range of backgrounds that have an $SU(2)$-isometry, there must be much more of interest that can be generated in a similar way to what is shown here. It would however be desirable to have some general understanding of the effect on the gauge theories before one actually performs the dualisation, like one does for $g$-structure rotation. It would be interesting to add flavours to the dual background considered here in the spirit of \cite{Barranco:2013fza}, where it is shown that one may flavour the dual solution by simply dualising the original solution with smeared flavours added. This would surely work and it would be interesting to see the effect this had on the Chern-Simons levels that are defined here. Flavour is added to the Maldacena-Nastase solution in \cite{Macpherson:2012zy,Canoura:2008at}. It would also be interesting to dualise the $G_2$-structure rotated solution. The baryonic branch of Klebanov-Strassler requires a $B_2$ with a leg in the holographic coordinate $r$ to give the correct $H_3$, this term causes a strange mixing of $r$ with the internal space in its non-abelian T-dual geometry which deforms the UV \cite{Itsios:2013wd}. The $B_2$ of the $G_2$-structure rotated solution is defined over the compact part of the internal manifold only and so the dual geometry would maintain the $AdS_4$ asymptotics, although the internal space may no longer be UV finite. The field theory dual to this would likely have some interesting properties.

\section*{Acknowledgements}
I would like to thank Carlos N\'u\~nez for useful discussions during this project as well as J\'er\^ome Gaillard and Daniel Thompson for concurrent collaboration which has been instructive. My work is supported by an STFC studentship.
\providecommand{\href}[2]{#2}\begingroup\raggedright\endgroup

\end{document}